\documentclass[conference]{IEEEtran}

\usepackage{amsmath,amssymb,amsfonts}
\usepackage{algorithmic}
\usepackage{graphicx}
\usepackage{textcomp}
\usepackage{xcolor}
\usepackage{algorithmic}
\usepackage[T1]{fontenc}
\usepackage{algorithm}

\usepackage{cases}
\usepackage{cite} 
\usepackage{amsthm}

\usepackage{subfigure}
\usepackage{amsmath}
\usepackage{enumitem}
\allowdisplaybreaks  

\def\BibTeX{{\rm B\kern-.05em{\sc i\kern-.025em b}\kern-.08em T\kern-.1667em\lower.7ex\hbox{E}\kern-.125emX}}

\begin{document}

\title{Knowledge-Guided Attention-Inspired Learning for Task Offloading in Vehicle Edge Computing}
\author{\IEEEauthorblockN{Ke Ma}
\IEEEauthorblockA{\textit{ Department of Electrical and Computer
Engineering} \\
\textit{University of California, San Diego and San Diego State
University}\\
La Lolla, USA \\
kem006@ucsd.edu}
\and
\IEEEauthorblockN{Junfei Xie}
\IEEEauthorblockA{\textit{ Department of Electrical and Computer
Engineering} \\
\textit{San Diego State University}\\
 San Diego, USA \\
jxie4@sdsu.edu}
}

\maketitle
\begin{abstract}
Vehicle edge computing (VEC) brings abundant computing resources close to vehicles by deploying them at roadside units (RSUs) or base stations, thereby enabling diverse computation-intensive and delay-sensitive applications. 
Existing task offloading strategies are often computationally expensive to execute or generate suboptimal solutions. In this paper, we propose a novel learning-based approach, Knowledge-guided Attention-inspired Task Offloading (KATO), designed to efficiently offload tasks from moving vehicles to nearby RSUs. KATO integrates an attention-inspired encoder-decoder model for selecting a subset of RSUs that can reduce overall task processing time, along with an efficient iterative algorithm for computing optimal task allocation among the selected RSUs. Simulation results demonstrate that KATO achieves optimal or near-optimal performance with significantly lower computational overhead and generalizes well across networks of varying sizes and configurations.


\end{abstract}


\section{Introduction}

The growing intelligence of modern vehicles demands greater computation resources to efficiently process the vast volumes of data generated by onboard sensors. Traditional cloud computing approaches often fall short due to latency and bandwidth constraints. In recent years, vehicular edge computing (VEC) emerges as a promising solution by enabling the offloading of computation-intensive and delay-sensitive tasks from vehicles to nearby edge servers, which are typically deployed at roadside units (RSUs) or base stations \cite{liu2021vehicular}. 

The problem of how to offload computation tasks from vehicles to nearby RSUs has been extensively studied \cite{jiang2025vehicle}. Recognizing that neighboring vehicles may have idle resources to share, other works \cite{huang2024joint} have explored task offloading from vehicles to nearby moving/parking vehicles. To fully utilize all available resources, studies such as \cite{yin2024joint} have investigated hybrid offloading scenarios in which both vehicle-to-RSU (V2R) and vehicle-to-vehicle (V2V) options are available.

Existing task offloading approaches can be broadly categorized into four groups: numerical optimization-based methods, heuristic methods, game-theoretic approaches, and learning-based techniques. Numerical optimization-based methods \cite{liu2022mobility, zhang2024computation} formulate the offloading problem as a mathematical optimization model and typically rely on commercial solvers such as Gurobi \cite{shen2025integrated} to compute solutions. For example, \cite{zhang2024computation} considers both task allocation and bandwidth resource allocation, formulating the problem as a fractional programming task aimed at maximizing computation efficiency. To solve it, the problem is decoupled into task allocation and bandwidth allocation subproblems and solved iteratively using an alternating optimization strategy. 
While these methods can achieve optimal or near-optimal results, they are often computationally expensive and are impractical for large-scale or real-time scenarios. 

Heuristic methods \cite{jiang2025vehicle, abuthahir2024tasks} are generally faster and easier to implement by applying approximations or rule-based strategies that reduce computational complexity. For example, \cite{jiang2025vehicle} applies the bat algorithm to jointly optimize service caching, content retrieval, and task offloading. Although heuristic algorithms offer scalable solutions with low computational overhead, they typically lack optimality guarantees and require expert knowledge to design effective problem-specific rules.

Game-theoretic approaches \cite{huang2024joint, yin2024joint} formulate the task offloading problem as a strategic game, in which vehicles and RSUs interact to optimize their individual utilities defined according to the task offloading objective. For example, \cite{huang2024joint} introduces a coalition formation game to jointly optimize spectrum sharing and task offloading. Although the game-theoretic formulations can effectively model distributed decision-making and complex interdependencies, they often require numerous iterations to converge, and the resulting Nash equilibrium may not be unique or easy to reach, particularly in dynamic or large-scale environments. 

The rapid advancement of machine learning has motivated researchers to develop learning-based techniques \cite{zhao2024incentive, lin2025multi} for deriving optimal task offloading decisions/policies from data. Among these, reinforcement learning (RL) approaches, such as the Proximal Policy Optimization (PPO), are commonly used. For example, \cite{lin2025multi} formulated the task offloading problem as a decentralized partially observable Markov decision process and proposed both non-cooperative and cooperative multi-agent PPO-based solutions. While these RL-based methods are model-free and well-suited for adapting to dynamic environments, they lack optimality guarantees and often suffer from convergence instability.

Given the dynamic and uncertain nature of vehicular networks, primarily caused by vehicle mobility, there is a crtical need for task offloading methods that can rapidly generate high-quality solutions and adapt effectively to changes in network topology and scale. To address this, we propose a novel learning-based approach for V2R task offloading that efficiently produces optimal or near-optimal solutions.  
Our approach, named Knowledge-guided Attention-inspired Task Offloading (KATO), tackles the offloading problem in two stages. The first stage selects a subset of RSUs using a learning-based selection strategy built on a knowledge-guided, attention-inspired encoder-decoder architecture. 
The second stage determines the optimal task allocation among the selected RSUs and the vehicle to minimize overall task completion time through an efficient iterative algorithm.
Simulation results demonstrate that KATO achieves solution quality comparable to the Gurobi-based optimal method while significantly reducing computational overhead and scaling effectively with network size. It also significantly outperforms heuristic and learning-based baselines of similar complexity in terms of solution quality. Moroever, it demonstrates strong generalization to smaller networks. When trained on large networks, this 
property makes KATO highly adaptable to dynamic environments and diverse VEC scenarios.

The rest of the paper is organized as follows. In Sec. \ref{sec:modeling}, we present system models and formulate the task offloading problem to be solved. KATO is then described in Sec. \ref{sec: method}, followed by simulation studies in Sec. \ref{sec:simulation}. 
Finally, we conclude the paper and discuss future works in Sec. \ref{sec:conclusion}.

\section{System Modeling and Problem Formulation} \label{sec:modeling}
\begin{figure}
    \centering
    \includegraphics[width=0.54\linewidth]{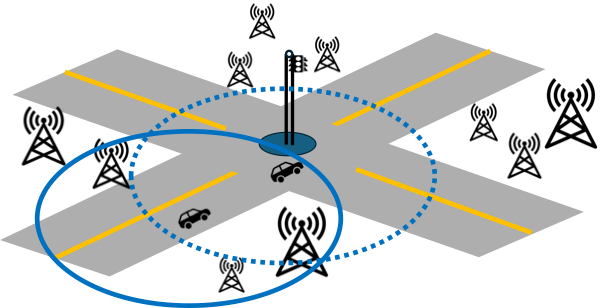}
    \caption{Illustration of the VEC system, where blue circles indicate the communication range of the moving vehicle at different positions along its trajectory.}
    \label{fig:Systemoverview}
    \vspace{-0.3cm}
\end{figure}

Consider a scenario (Fig.~\ref{fig:Systemoverview}) in which a vehicle, indexed as $0$, is moving along a straight road at a constant speed of $v$ (m/s). 
Suppose the vehicle has a computation-intense task to process, characterized by the pair $(Q, T)$, where $Q$ represents the task size 
and $T$ is the deadline before which the task must be completed.  To accelerate processing and reduce energy consumption, the vehicle offloads the task to nearby RSUs equipped with computational resources. Assume that a set of $n-1$ RSUs, denoted by $\mathcal{S} = \{1,...,n-1\}$, are currently within the communication range (denoted as $\xi >0$) of this vehicle. Let $\mathcal{N} = \{0\} \cup \mathcal{S}$ denote the set comprising both the vehicle and all RSUs. 

Assuming the task is arbitrarily decomposable \cite{ma2024jointtaskallocationscheduling}, the vehicle seeks to optimally partition the task and offload task portions to these RSUs. However, 
due to the vehicle's mobility, some RSUs may fall outside its communication range over time, resulting in failures to return the computed results and causing the entire task to fail. Moreover, as shown in our previous study \cite{ma2024decentralized}, increasing the number of compute nodes does not necessarily reduce the overall computation time. Therefore, careful selection of RSUs is essential to ensure successful task completion while minimizing the total task processing time.

\subsection{System Modeling}

To model  data transmission between RSUs and the vehicle, we adopt the approach described in \cite{ma2024decentralized}.  
Specifically, assume that the RSUs communicate with the vehicle using the Orthogonal Frequency Division Multiplexing protocol \cite{10622383}. 
When the vehicle transmits data to $m$ RSUs ($m\leq n-1$) simultaneously, the available bandwidth is equally divided among them, similar bandwidth partition strategy was also used in \cite{huo2024efficient}. The data transmission rate for the link between the vehicle and the $i$-th RSU is given by
$R_{i}(t) = \frac{B}{m}log_2(1 
 +\frac{\eta_{i}}{d_{i}(t)^{2}})$, where $B$ (MHz) denotes the total available bandwidth, $\eta_{i}$ represents the signal-to-noise (SNR) ratio between RSU $i$ and the vehicle, and $d_i(t)$ is the distance between them at time $t$.

To model task processing, we use the computation model from \cite{10622383}. Let $b$ represent the number of CPU cycles required to compute 1 bit of data, and $f_i$ (GHz) denote the computation capacity of RSU $i$. Given a task of size $q$ (Gbits), the time required by RSU $i$ to process the task  is $ T_i^{comp} = q \beta_i$, 
where $\beta_i = \frac{b}{f_i}$ (s/Gbit) indicates the time taken by RSU $i$ to process one Gbit of data.

To model the movement of the vehicle, we define its heading direction, which aligns with the road, as the $x$ axis. The initial location of the vehicle before executing the task is denoted by $(x_0, y_0)$. The location of each RSU $i \in \mathcal{S}$ is given by $(x_i, y_i), i \in \mathcal{S}$, which is fixed. 
The distance between the vehicle and each RSU $i$ after time $t$ is calculated as:
\begin{align}
    d_i(t) = \sqrt{(x_i - (x_0 + vt))^2 + (y_i - y_0)^2} . \label{eq:dis}
\end{align}

\subsection{Problem Formulation}
The vehicle selects proper compute nodes from among all RSUs and itself.  To represent its selection, we introduce binary decision variables $z_i$ for each $i\in \mathcal{N}$, where $z_i =1$ if node $i$ is selected to process a portion of the task, and $z_i =0$  otherwise. The total number of selected nodes is denoted by $m = \sum_{i \in \mathcal{S}}z_i$. 
Additionally, we introduce continuous decision variables $q_i \in \mathbb{R}_{\geq 0}$ to represent the size of task portion assigned to each node $i\in \mathcal{N}$. The transmission time for each node to receive its assigned subtask can then be expressed as
  $  T_{i}^{tran} = \begin{cases}
    \frac{z_iq_i}{R_i(0)}, &\text{~if~} i\in\mathcal{S} \\
    0, &\text{~if~} i=0 \end{cases}$. 
The total time required for each node $i$ to receive and process its assigned task is given by  
    $U_i = T_i^{tran} + T_i^{comp}$, 
where $T_i^{comp} = q_i \beta_i$. 
Here, we assume that the generated results are small in size, making the transmission delay for sending them back to the vehicle negligible, as is often assumed in existing studies \cite{ma2024decentralized}. 

The objective of this study is to minimize the total task completion time by optimizing the node selection and task allocation formulated as follows: 
\begin{align*}
    \mathcal{P}: & \min_{q_i, z_i, i \in \mathcal{N}} \quad \max_{i \in \mathcal{N}} U_i \\
    s.t. \quad & \sum_{i \in \mathcal{N}} z_iq_i = Q & C_1 \\
    & z_i \in \{0,1\}, 0 \leq q_i \leq Q, \forall i\in \mathcal{N} &C_2\\
    & \sum_{i \in \mathcal{N}} z_i \geq 1 & C_3\\ 
   & U_i \leq T, \forall i\in \mathcal{N} & C_4 \\
    & z_i d_i(U_i) \leq \xi, \forall i \in \mathcal{S} & C_5
\end{align*}
In the above formulation, constraint $C_1$ ensures the task size integrity. Constraints $C_2$ enforce that the control variables take valid values. 
Constraint $C_3$ gaurantees that the task is assigned to at least one node. Constraints $C_4$ ensure the task is completed before the specified deadline. 
The last constraint $C_5$ requires that the selected RSUs be within the vehicle's communication range at the time they finish processing the assigned subtask, so that the results can be successfully transmitted back to the vehicle. 
This results in a complex combinatorial mixed-integer nonlinear optimization problem, which cannot be directly solved. Moreover, the search space grows exponentially with the size of the network ($n$). 

\section{Knowledge-Guided Attention-Inspired Task Offloading}
\label{sec: method}
\begin{figure}
        \centering
        \includegraphics[width=0.9\linewidth]{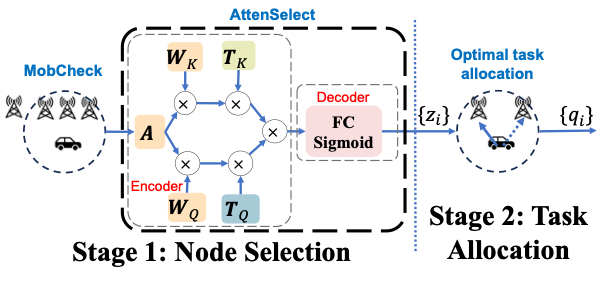}
        \caption{System architecture of KATO. }
        \label{fig:Transformer}
    \end{figure}

In this section, we present the proposed approach, KATO, to efficiently solve problem $\mathcal{P}$. KATO solves this problem in two stages: first, we optimize node selection by determining $z_i$, $\forall i \in \mathcal{N}$; then, we optimize task allocation by determining $q_i$, $\forall i \in \mathcal{N}$, as illustrated in Fig. \ref{fig:Transformer}.

\subsubsection{Node Selection}
The node selection stage faces two key challenges. Firstly, due to the vehicle's movement, RSUs that are initially within communication range may fall out of range over time, potentially leading to failed result delivery if tasks are assigned to them. Secondly, there is a tradeoff between utilizing additional computing resources by selecting more RSUs and the increased transmission delays caused by bandwidth sharing among a larger number of RSUs. The node selection stage addresses the first challenge through a \textit{MobCheck} module, which predicts the distance between the vehicle and each RSU $i\in\mathcal{S}$ after time $\min \{T, T_{\text{loc}}\}$ using \eqref{eq:dis}, where $T_{\text{loc}} = T_0^{comp} = Q\beta_0$ denotes the time required by the vehicle to process the entire task locally, representing the worst-case scenario. RSUs with predicted distances exceeding the vehicle's communication range ($\xi$) are then filtered out, ensuring that all remaining RSUs will be within range upon task completion. These remaining RSUs, denoted as $\mathcal{S}'$, are subsequently evaluated using the \textit{AttenSelect} module for final selection, which is designed to address the second challenge.

The \textit{AttenSelect} module selects nodes by analyzing their computing and communication characteristics to assess the merit of each node based on its potential to speed up computation. However, due to shared bandwidth constraints, an RSU’s transmission delay is affected by the total number and characteristics of other RSUs involved in offloading. This makes the contribution of each RSU inherently dependent on the others.  To capture such interdependencies, the attention mechanism is particularly well-suited. Attention-based models, such as Transformers \cite{vaswani2017attention}, process the input as a sequence of data points and interpret each element in the context of the entire sequence. The attention mechanism assigns scores to individual elements, which capture their relative importance in computing the output. Inspired by this mechanism, we design a knowledge-guided, attention-like model to capture inter-node dependencies and compute their merit scores. 

Denote $\mathcal{N}' = \{0\} \cup \mathcal{S}'$, and $n' = |\mathcal{N}'|$.
The input to the encoder, denoted $\boldsymbol{A}=[\boldsymbol{a}_0; \boldsymbol{a}_1; \ldots, \boldsymbol{a}_{n'-1}]\in \mathbb{R}^{n' \times 4}$, encapsulates the key features of all remaining RSUs as well as the vehicle. It is treated as a sequence of $n'$ elements, each corresponding to a compute node and represented by a four-dimensional feature vector $\boldsymbol{a}_i = [x_i, y_i, \beta_i, \eta_{i}]$, $i\in \{0,1,\ldots, n'-1\}$, representing their position $(x_i, y_i)$ (before task execution), computation power $\beta_i$ and the SNR ratio $\eta_{i}$. Since the time required for transmitting results back to the vehicle is negligible, the total transmission delay is determined solely by the offloading time needed to send task data from the vehicle to the RSUs. Therefore, $\boldsymbol{A}$ contains all the information necessary to determine the impact of each node on the task processing time. 

Rather than directly applying a traditional attention mechanism to compute merit scores, we design our encoder based on the intuition that nodes with greater computing power are more likely to be selected. 
To capture this, the encoder is structured such that a node's merit reflects the delay reduction it contributes when added to a set of more powerful nodes that are assumed to have already been selected. This is achieved by pre-sorting the rows of $\boldsymbol{A}$ such that the first row corresponds to the vehicle (the offloader), with $\eta_{0}$ set to $0$. The remaining rows are arranged in descending order according to the RSUs' computing power $\beta_i$. 
The encoder then computes the key and query matrices as follows:
   \begin{align}
    \boldsymbol{K}  &= \boldsymbol{T}_K  \boldsymbol{A} \boldsymbol{W}_K
     = \begin{bmatrix}
        1 & 0 &\ldots & 0 \\
        1 & 1 &\ldots & 0 \\
        \vdots & \vdots & \ddots & \vdots \\
        1 & 1 & \ldots & 1
    \end{bmatrix} \begin{bmatrix}
         \boldsymbol{a}_0\\
         \boldsymbol{a}_1\\
        \vdots \\
        \boldsymbol{a}_{n'-1} 
    \end{bmatrix} \boldsymbol{W}_K\notag \\
    & = \begin{bmatrix}
        \boldsymbol{a}_0 \\
        \boldsymbol{a}_0 +   \boldsymbol{a}_1\\
        \vdots \\
        \boldsymbol{a}_0 + \boldsymbol{a}_1+\cdots + \boldsymbol{a}_{n'-1} 
    \end{bmatrix} \boldsymbol{W}_K\label{eq: t1}
    \end{align}
    \begin{align}
        \boldsymbol{Q} &= \boldsymbol{T}_Q  \boldsymbol{A} \boldsymbol{W}_Q
         = \begin{bmatrix}
            1 & 0 & \cdots & 0
        \end{bmatrix} \begin{bmatrix}
         \boldsymbol{a}_0\\  \boldsymbol{a}_1 \\
        \vdots \\
        \boldsymbol{a}_{n'-1} 
    \end{bmatrix} \boldsymbol{W}_Q = 
             \boldsymbol{a}_0
      \boldsymbol{W}_Q\label{eq: t2}
    \end{align}
where $\boldsymbol{W}_K, \boldsymbol{W}_Q\in \mathbb{R}^{4 \times h}$ are learnable weight matrices and $h$ denotes the dimensionality of the hidden feature space. Unlike the standard attention mechanism, we introduce a transformation matrix $\boldsymbol{T}_K\in \mathbb{R}^{n' \times n'}$ to aggregate the features of increasingly larger subsets of nodes with higher computing capabilities. 
Additionally, we introduce a transformation matrix $\boldsymbol{T}_Q \in \mathbb{R}^{1 \times n'}$ to ensure that the query focuses specifically on the case where the vehicle (node 0) is the offloader. 
Given the key and query matrices, the merit scores are computed as:
 \begin{align}
        \boldsymbol{e} = \frac{\boldsymbol{Q}\boldsymbol{K}^T}{\sqrt{h}} \label{eq:score}   
    \end{align}
which reflects the estimated contribution of each compute node to performance improvement when added to an already selected subset of more powerful nodes. Notably, the value matrix is omitted, as the objective is not to generate a weighted combination of values.

Based on the computed merit scores, the decoder generates the node selection decisions. It is implemented as a fully connected layer followed by a sigmoid activation function. Specifically, the decoder output is computed as:
    \begin{align}
        \boldsymbol{p} = \sigma (g(\boldsymbol{e})) \label{eq: transformereq}
    \end{align}
where $\sigma$ denotes the sigmoid function, and $g$ represents a fully connected layer. The resulting vector $\boldsymbol{p}$ represents selection probabilities for each node. A node is selected if its corresponding probability satisfies $p_i \geq 0.5$. That is $z_i = 1$ if $p_i \geq 0.5$, and $z_i = 0$ otherwise.

\subsubsection{Task Allocation}
Given the set of compute nodes selected by the \textit{AttenSelect} module, denoted as $\mathcal{N}''$, the task allocation stage aims to determine the optimal task allocation $q^*_i$ for each node $i\in \mathcal{N}''$. To compute $q^*_i$ efficiently, we adopt an iterative procedure derived from our previous work \cite{ma2024decentralized}.
In each iteration, a single RSU is randomly selected for assessment, and the task allocations are updated for the subset of compute nodes that have been assessed to that point. 
Each iteration consists of two steps. The first step calculates the processing capacity, denoted as $c_i$, for each node under consideration in the current iteration, assuming that only these nodes participate in task execution. The second step determines the optimal task allocation among these nodes. 
Specifically, let $\mathcal{F}^k \subseteq  \mathcal{S}'': = \mathcal{N}''\setminus \{0\}$ denote the set of RSUs considered in the $k$-th iteration, where the superscript $k$ indicates the iteration index. Notably, the vehicle (node $0$) is always included in task execution, as its involvement will contribute to reducing the overall task processing time. 
Initially ($k=0$), $\mathcal{F}^0 = \emptyset$, and only the vehicle is considered for task processing, with its processing capacity denoted as $c^0_0 = \beta_0$. The resulting optimal task allocation for the vehicle is $q^0_0 = Q$, corresponding to the case of local computing.  

In the next iteration ($k=1$), an RSU, denoted as $s^1 \in \mathcal{S}''$, is randomly picked from the set $\mathcal{S}''$ for assessment and added to the set by $\mathcal{F}^1 \leftarrow \mathcal{F}^0\cup \{s^1\}$. The processing capacity of RSU $s^1$ is computed as 
$c^1_{s^1} = \beta_{s^1} + \frac{1}{Blog_2(1+\eta_{s^1}d_{s^1}^{-2})}$, which represents the time required for it to receive and process one bit of data. This corresponds to the case when only the vehicle and this RSU $s^1$ participate in computation, and RSU is allocated the entire bandwidth. The vehicle's processing capacity remains unchanged, i.e., $c^1_0 = \beta_0$. The optimal task allocation can then be derived as $q^1_0 = \frac{c^1_{s^1} Q}{c^1_0 + c^1_{s^1}}$ and $q^1_{s^1} = \frac{c^1_0Q}{c^1_0 + c^1_{s^1}}$, according to Lemma 1 in \cite{ma2024decentralized}. 

In each of the following iterations ($1<k \leq |\mathcal{S}''|$), one RSU $s^k$ is picked without replacement from $\mathcal{S}''$ and added to the set $\mathcal{F}^k$. For each RSU $i$ in $\mathcal{F}^k$, its processing capacity is updated as $c_i^k = \beta_i + \frac{|\mathcal{F}^k|}{Blog_2(1+\eta_id_i^{-2})}$, as the available bandwidth is now shared among a larger number of participating RSUs, and $c_0^k = \beta_0$. The optimal task allocation is computed by the following equations \cite{ma2024decentralized}: 
\begin{align}
q_i^k & = \sum_{j \in \mathcal{F}^{k-1} \cup \{0\}}(1 - \frac{c_j^{k-1}}{\mathtt{I}^k c_j^k})q_j^{k-1} \label{eq: yi bar}\\
\mathtt{I}^k &= \frac{q_{s^k}^{k-1}c_{s^k}^{k-1}}{q_{s^k}^{k}c_{s^k}^k} \label{eq: yj first}
\end{align}

This process continues until all selected RSUs have been evaluated, i.e., when $k= |\mathcal{S}''|$. 
Algorithm \ref{Algo: Cluster Member Selection} summarizes the complete procedure of the proposed approach.

\begin{algorithm}
    \caption{KATO} \label{Algo: Cluster Member Selection}
    \begin{algorithmic}[1]
        \STATE Find $\mathcal{S}'$ by filtering out RSUs in $\mathcal{S}$ that fall out of the vehicle's communication range at time $\min \{T, T_{\text{loc}}\}$; 
        \STATE Construct matrix $\boldsymbol{A}$ using information about the remaining RSUs and the vehicle;
        \STATE Feed $\boldsymbol{A}$ into the \textit{AttenSelect} module to obtain $\mathcal{S}''$; 
        \STATE Determine the optimal task allocation for the RSUs in $\mathcal{S}''$ and the vehicle by following the above iterative procedure;
    \end{algorithmic}
\end{algorithm}

\section{Simulation studies} \label{sec:simulation}
In this section, we conduct simulation studies to evaluate the performance of our approach.

\subsection{Experiment Setup}
We randomly generate and distribute RSUs within a $100\times100$m area. The communication range of the moving vehicle $\xi$ is set to 150m. 
The computing power of every node $f_i, i \in \mathcal{N}$ is sampled from a uniform distribution over the range $[0.1, 10]$ GHz, with $b=1$. The SNR ratio $\eta_{i}$ is sampled from a uniform distribution over the range $[20, 30]$dbm. The task size $Q$ is set to $1$Gbits, and the bandwidth $B$ is set to 50GHz. $h$ is set to 5. To train the \textit{AttenSelect} module, we use the Gurobi solver to generate labeled data by solving problem $\mathcal{P}$. A total of $100,000$ data samples are generated and subsequently divided into training, validation, and test sets using a 70\%/15\%/15\% split. The model is trained using the binary cross-entropy loss function and the Adam optimizer. We set the number of training epochs to 20 and use a learning rate of 0.001.

\subsection{Experiment Results}
\subsubsection{Mobility Management} 
\begin{figure}
    \centering
    \includegraphics[width=0.5\linewidth]{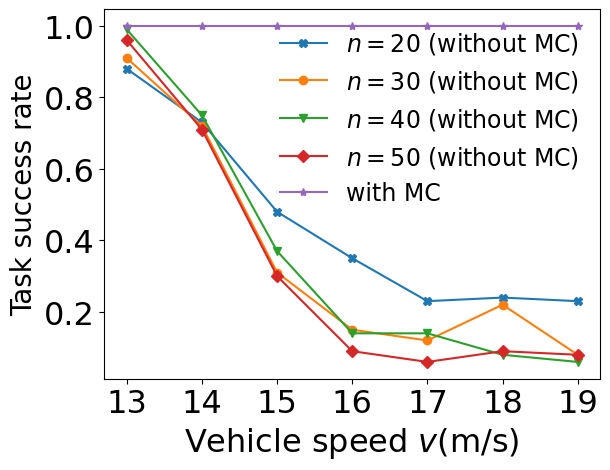}
    \caption{Comparison of task success rate of our approach with and without the \textit{MobCheck} (MC) module under varying vehicle speeds and network sizes.}
    \label{fig:RSU_successRate}
    \vspace{-0.3cm}
\end{figure}

We first evaluate the effectiveness of the \textit{MobCheck} module in the node selection stage in addressing potential task failures caused by the vehicle's movement. 
We vary both the total number of RSUs originally within the vehicle's communication range and the vehicle's mobility by adjusting its speed from 13m/s to 19m/s. For each combination of network size and speed setting, we randomly select 100 samples from the test set and measure the task success rate. 
Fig. \ref{fig:RSU_successRate} shows the performance of our approach with and without the \textit{MobCheck} module. When mobility is taken into account, all tasks are completed successfully, demonstrating the effectiveness of the \textit{MobCheck} module. In constrast, when mobility is not considered, 
the task success rate decreases dramatically as the vehicle speed increases, as the selected RSUs are more likely to fall out of the communication range. 

\subsubsection{Optimality and Efficiency}
To evaluate the optimality and efficiency of our approach, we compare it with the following baselines:
\begin{itemize}
    \item \textbf{DECC}\cite{10521565}: This method implements a partial offloading approach. The task is partitioned and offloaded to the RSU which minimizes the task processing delay.
    \item \textbf{Load Balance (LB)}\cite{jin2024task}: This method partitions the task and distributes it across all RSUs in proportion to their computing power. 
    \item \textbf{Multilayer Perceptron (MLP)}: This method adopts an MLP \cite{gardner1998artificial} to select nodes. To maintain a comparable level of complexity with our approach, we configure the MLP with a single hidden layer consisting of five units.
    \item \textbf{Standard Attention (SA)}: This method adopts a standard attention mechanism \cite{vaswani2017attention} for node selection, without incorporating the transformation matrices $\boldsymbol{T}_K$ and $\boldsymbol{T}_Q$ to select nodes. Similar to our approach, it also omits the use of a value matrix.
    \item \textbf{Optimal}: This method employs Gurobi \cite{shen2025integrated} to compute the optimal node selection given a set of compute nodes. 
\end{itemize}
The \textit{MobCheck} module is incorporated into all baseline methods to account for vehicle mobility, with  $v=15 (m/s)$. The \textbf{MLP}, \textbf{SA} and \textbf{Optimal} all adopt the same task allocation strategy used in our approach, which computes the optimal task allocation given a set of selected compute nodes. 

\begin{figure}
  \centering
  \subfigure[]{
    \centering\includegraphics[width=0.45\linewidth]{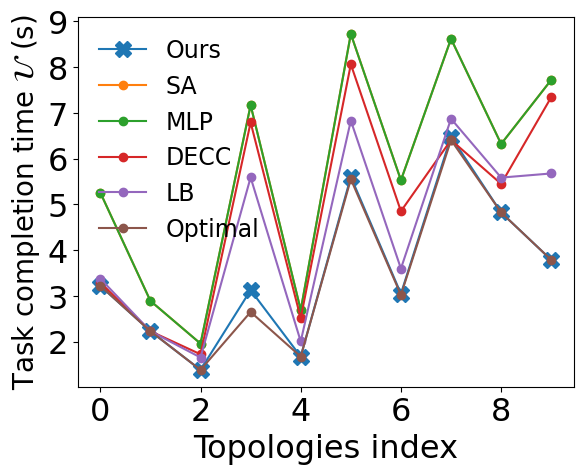}
    \label{fig:RSU_resule100R150N19}}
  \subfigure[]{
    \centering\includegraphics[width=0.45\linewidth]{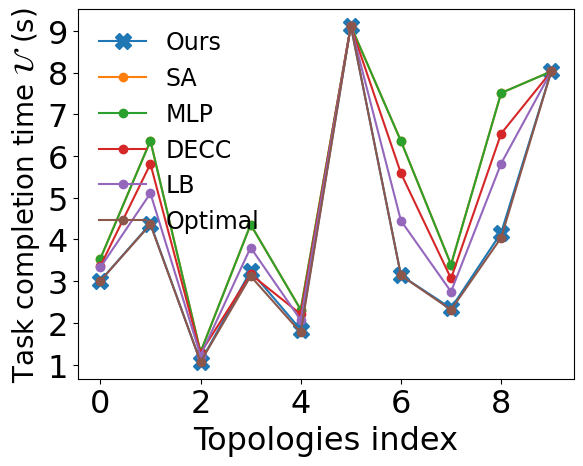}
    \label{fig:RSU_resule100R150N29}}
  \subfigure[]{
    \centering\includegraphics[width=0.45\linewidth]{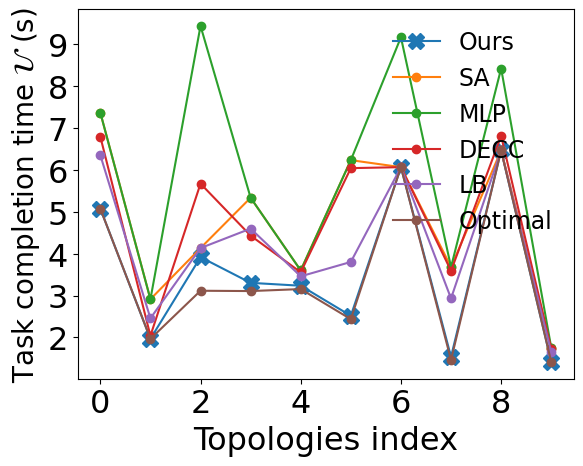}
    \label{fig:RSU_resule100R150N39}}
  \subfigure[]{
    \centering\includegraphics[width=0.45\linewidth]{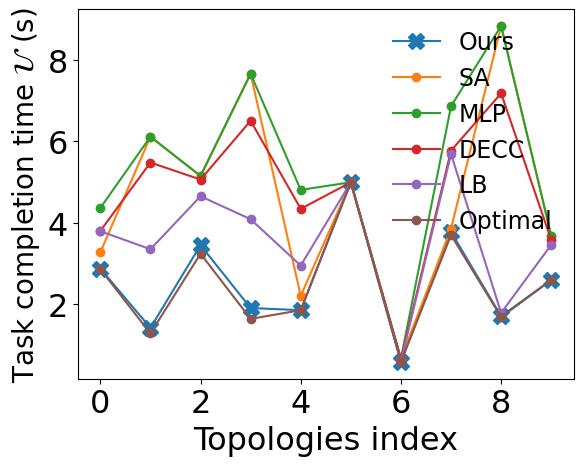}
    \label{fig:RSU_resule100R150N49}}
\caption{Performance comparison of different approaches across various network topologies with (a) $n=20$, (b) $n=30$, (c) $n=40$ and (d) $n=50$ nodes.} \label{fig: OffloadingMobility}
\vspace{-0.3cm}
\end{figure}

For a comprehensive assessment, we vary network size $n$, and randomly generate 10 different topologies for each size. Fig. \ref{fig: OffloadingMobility} shows the comparison results with \textbf{DECC}, \textbf{LB}, \textbf{MLP} and \textbf{SA}. As we can see, our approach generates optimal or near-optimal solutions and significantly outperforms all heuristic or learning-based baselines. Moreover, the comparison with \textbf{SA}, the standard attention-based approach, demonstrates the effectiveness of our design. 

\begin{figure}
  \centering
  \subfigure[]{
    \centering\includegraphics[width=0.45\linewidth]{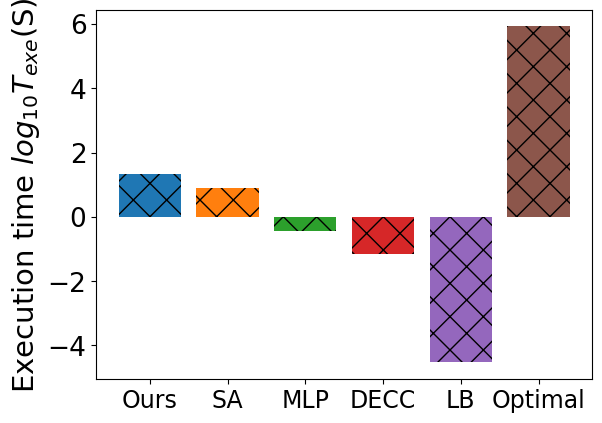}
    \label{fig:RSU_result_time100R150N19}}
  \subfigure[]{
    \centering\includegraphics[width=0.45\linewidth]{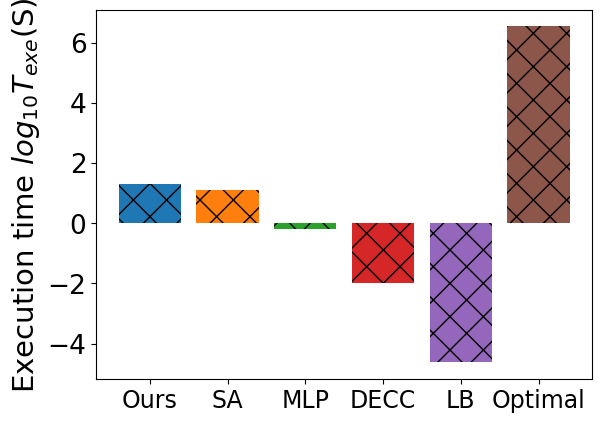}
    \label{fig:RSU_result_time100R150N19}}
  \subfigure[]{
    \centering\includegraphics[width=0.45\linewidth]{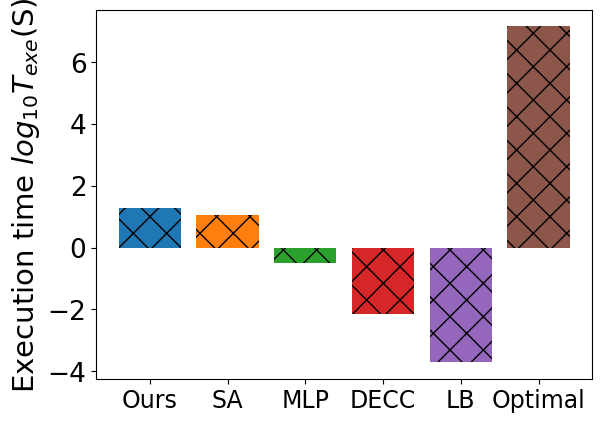}
    \label{fig:RSU_result_time100R150N39}}
  \subfigure[]{
    \centering\includegraphics[width=0.45\linewidth]{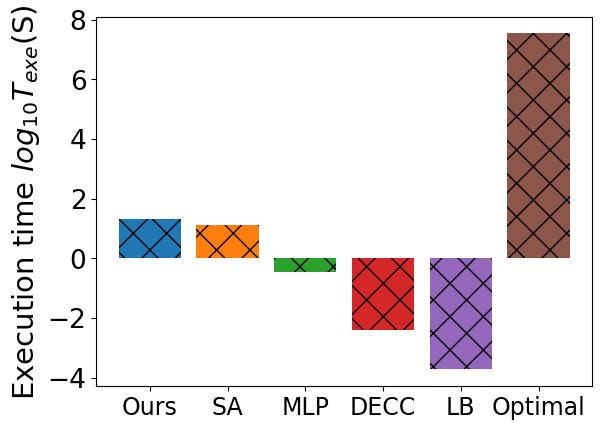}
    \label{fig:RSU_result_time100R150N49}}
\caption{Average execution time of different approaches when (a) $n=20$, (b) $n=30$, (c) $n=40$ and (d) $n=50$.} \label{fig: TransformerSingleSame_time}
\vspace{-0.3cm}
\end{figure}

Fig. \ref{fig: TransformerSingleSame_time} shows the average execution time of different methods. While our approach is less efficient than the heuristic or learning-based baselines, it significantly outperforms the \textbf{Optimal} method in terms of runtime, achieving a good balance between solution quality and computational efficiency. 


\subsubsection{Generalization to Smaller Networks}
\begin{figure}
  \centering
  \subfigure[]{
    \centering\includegraphics[width=0.45\linewidth]{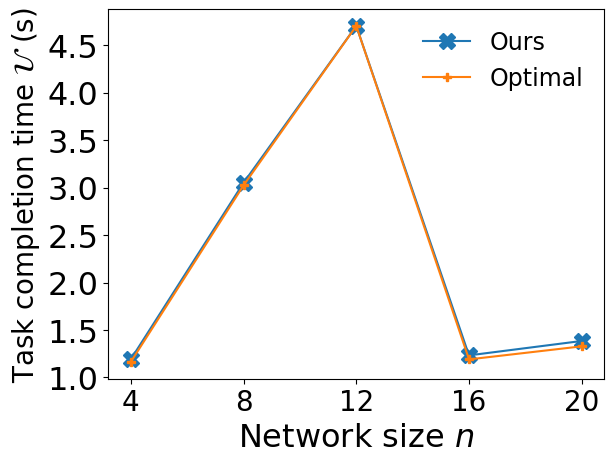}
    \label{fig:Transformer_different100R150N19}}
  \subfigure[]{
    \centering\includegraphics[width=0.45\linewidth]{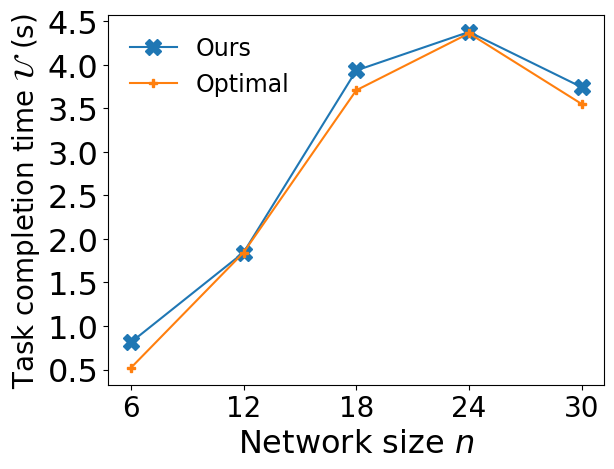}
    \label{fig:Transformer_different100R150N29}}
  \subfigure[]{
    \centering\includegraphics[width=0.45\linewidth]{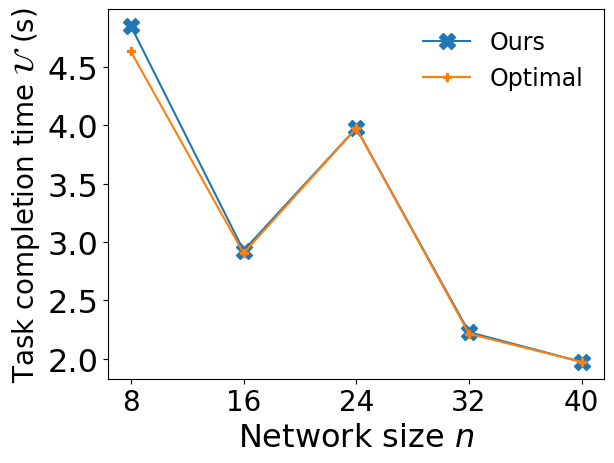}
    \label{fig:Transformer_different100R150N39}}
  \subfigure[]{
    \centering\includegraphics[width=0.45\linewidth]{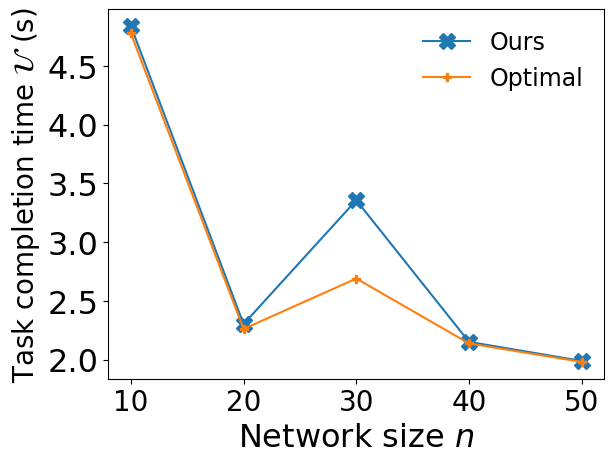}
    \label{fig:Transformer_different100R150N49}}

\caption{Performance of our approach across various smaller test networks, with models trained on networks containing (a) $20$, (b) $30$, (c) $40$ and (d) $50$ nodes.} \label{fig: TransformerSingleDifferent}
\vspace{-0.4cm}
\end{figure}

Our approach is input-size agnostic and generalizes well to networks smaller than those in the training set. To demonstrate this property, we generate smaller networks for testing by randomly removing RSUs from the original test set. Mobility is not considered in this setting. Fig. \ref{fig: TransformerSingleDifferent} shows the quality of the solutions generated by our approach, compared to the \textbf{Optimal} method across various network sizes. As shown, our approach still generates near-optimal solutions, demonstrating its strong generalization to smaller networks. This is an especially valuable property in real-world scenarios where network sizes often vary due to vehicle mobility.


\section{Conclusion and Future Works} \label{sec:conclusion}
In this paper, we introduced KATO, a novel learning-based approach for efficient task offloading in VEC. KATO combines an attention-inspired encoder-decoder model for context-aware compute node selection with an efficient iterative algorithm for optimal task allocation. To evaluate its performance, we conducted extensive comparative evaluations. Results demonstrate that KATO achieves a good balance between solution quality and computational efficiency, achieving performance comparable to optimal solutions while significantly reducing computational cost. Additionally, KATO exhibits strong scalability with respect to network size and generalizes well to smaller networks, making it highly adaptable to dynamic environments. In the future, we will extend our approach to incorporate V2V communication.


\vspace{-0.1cm}

\section*{Acknowledgment}
We would like to thank the National Science Foundation under Grants CAREER-2048266, CCRI-1730675, and CCF-2402689 for the support of this work.

\bibliographystyle{IEEEtran}
\bibliography{reference}

\end{document}